\pdfoutput=1 

\documentclass[12pt]{article}
\usepackage{graphicx}
\usepackage{geometry} 
\title{The PLANCK LFI flight model ortho-mode transducers}
\author{O. D'Arcangelo$^1$, A. Simonetto$^1$, L. Figini$^1$,E. Pagana$^2$, \\F.Villa$^3$, M. Pecora$^4$, P. Battaglia$^4$, M. Bersanelli$^5$, R.~C. Butler$^3$,\\S. Garavaglia$^1$, P. Guzzi$^4$, N. Mandolesi$^3$ and C. Sozzi$^1$ }
\date{} 
\begin{document}

\maketitle
\begin{center}
$^1$\emph{Istituto di Fisica del Plasma - CNR, via Cozzi 53, 20125 Milano, Italy} \\ 
$^2$\emph{Independant consultant}\\
$^3$\emph{Istituto di Astrofisica Spaziale e Fisica Cosmica, INAF, via P. Gobetti, 101, I40129 Bologna, Italy}\\
$^4$\emph{Thales Alenia Space Italia, s.s. Padana Superiore 290, 20090 Vimodrone (MI), Italy}\\
$^5$\emph{Universit\`a degli Studi di Milano, Via Celoria 16, 20133 Milano, Italy}
\end{center}
\begin{center} \textbf{Abstract}
\end{center}
The Low Frequency Instrument (LFI) of the ESA Planck CMB mission is an array of 22 ultra sensitive pseudocorrelation radiometers 
   working at $30$, $44$, and $70$ GHz. LFI has been calibrated and delivered for integration with the satellite to the European Space 
   Agency on November 2006. The aim of Planck is to measure the anisotropy and polarization of the Cosmic Background Radiation with 
   a sensitivity and angular resolution never reached before over the full sky. LFI is intrinsically sensitive to polarization thanks to the use of Ortho-Mode Transducers (OMT) located between the feedhorns and the 
   pseudo-correlation radiometers. The OMTs are microwave passive components that divide the incoming radiation into two 
   linear orthogonal components. A set of 11 OMTs (2 at 30 GHz, 3 at 44 GHz, and 6 at 70 GHz) were produced and tested.
   This work describes the design, development and performance of the eleven Flight Model OMTs of LFI. The final design was reached after several years of development. At first, Elegant Bread Board OMTs were produced 
   to investigate the manufacturing technology and design requirements. Then, a set of 3 Qualification Model (QM) OMTs were 
   designed, manufactured and tested in order to freeze the design and the manufacturing technology for the flight units. Finally, the 
   Flight Models were produced and tested. It is shown that all the OMT units have been accepted for flight and the electromagnetic performance is at least marginally compliant with 
   the requirements. Mechanically, the units passed all the thermoelastic qualification tests after a reworking necessary after the QM campaign.
\section{Introduction}
The PLANCK satellite, ESA's third generation space mission devoted to the study of the Cosmic Microwave Background (CMB), 
is designed to produce a map of the CMB anisotropy over the whole sky, with unprecedented combination of
angular resolution (4'--30') and sensitivity  ($\Delta T/T\simeq10^{-6}$), for a wide range of frequencies from 27 to 
850 GHz (Tauber, \cite{Tauber}). Two complementary instruments, the Low Frequency Instrument (LFI) (Bersanelli et al., \cite{bersanelli}) operating at three 20\% 
frequency bands centered at $30$, $44$ and $70$ GHz, and the High Frequency Instrument (HFI) working in the 100--850~GHz 
range, have been integrated together in the focal plane of a Gregorian off-axis 
optimized telescope (Fargant et al., \cite{Fargant}; Villa et al., \cite{villa}). 
LFI consists of an array of 11 corrugated feed horns (FH)  (Villa et al., \cite{villa1}), each connected through a dedicated Ortho-mode transducer 
(OMT) to a pair of ultra low noise pseudo-correlation receivers for a total of $44$ detector outputs (Seiffert 
et al., \cite{Seiffert}).
The OMT is a microwave passive component that splits the signal collected by the horn into two
linear orthogonally polarized components to be amplified and detected in the radiometer chain.
For the unpolarized sky component, the OMT acts as a
channel doubler, improving the sensitivity by a factor $1/\sqrt 2$. For the polarized sky component it will allow
disentanglement of the two orthogonal polarizations, transforming each pair of pseudo-correlation receivers into an $X-Y$ (X minus Y) polarimeter (Leahy et al., \cite{Leahy}). 
On the other hand, the OMT is a source of noise before the low noise amplifier, and also a critical component for what concerns the large bandwidth required,
so the theoretical benefit in sensitivity for the unpolarized component cannot be fully reached.

Due to the requirements imposed by electromagnetic performances and mechanical constraints, commercial components 
were unavailable. It was decided that the orthomode transducers should be designed and developed in the framework of the LFI 
industrial activity. 
The development of OMT components was carried out in three steps, following the development of the LFI 
radiometers. At first, prototypes were designed and manufactured with the aim of selecting the best design 
options and manufacturing technologies. Then, the Qualification Models (QM) were manufactured and 
tested to qualify the design, the manufacturing process and the testing procedure. Finally, the Flight Models (FM) and 
Flight Spares (FS) were built, tested and integrated on the LFI radiometer chain assemblies, ready 
to be launched. Small modifications, related to the external structure, were applied to the FM units compared with 
the QM, in order to reduce the spillover radiation entering the 4~K reference 
horns (Cuttaia et al., \cite{cuttaia}). 

This paper describes the development of the Ortho-mode transducers for the Planck LFI instrument. A scientific 
assessment on CMB polarization measurements with LFI is addressed in Leahy et al., \cite{Leahy}.

Section \ref{principles} describes the general principles of the OMT in order to describe the general design 
requirements assumed for LFI.
In section \ref{fabrication} the electromagnetic design and manufacturing technology are reported. In 
section \ref{tests}
the performance and verifications tests are described. The results obtained are shown in section \ref{results}, and their extrapolation to 
flight conditions is made in section \ref{flight}.  
The conclusions are addressed
in section \ref{conclusions}.

\section{OMT Principles and design requirements}\label{principles}
\begin{figure}
\centering
\includegraphics[width=6cm]{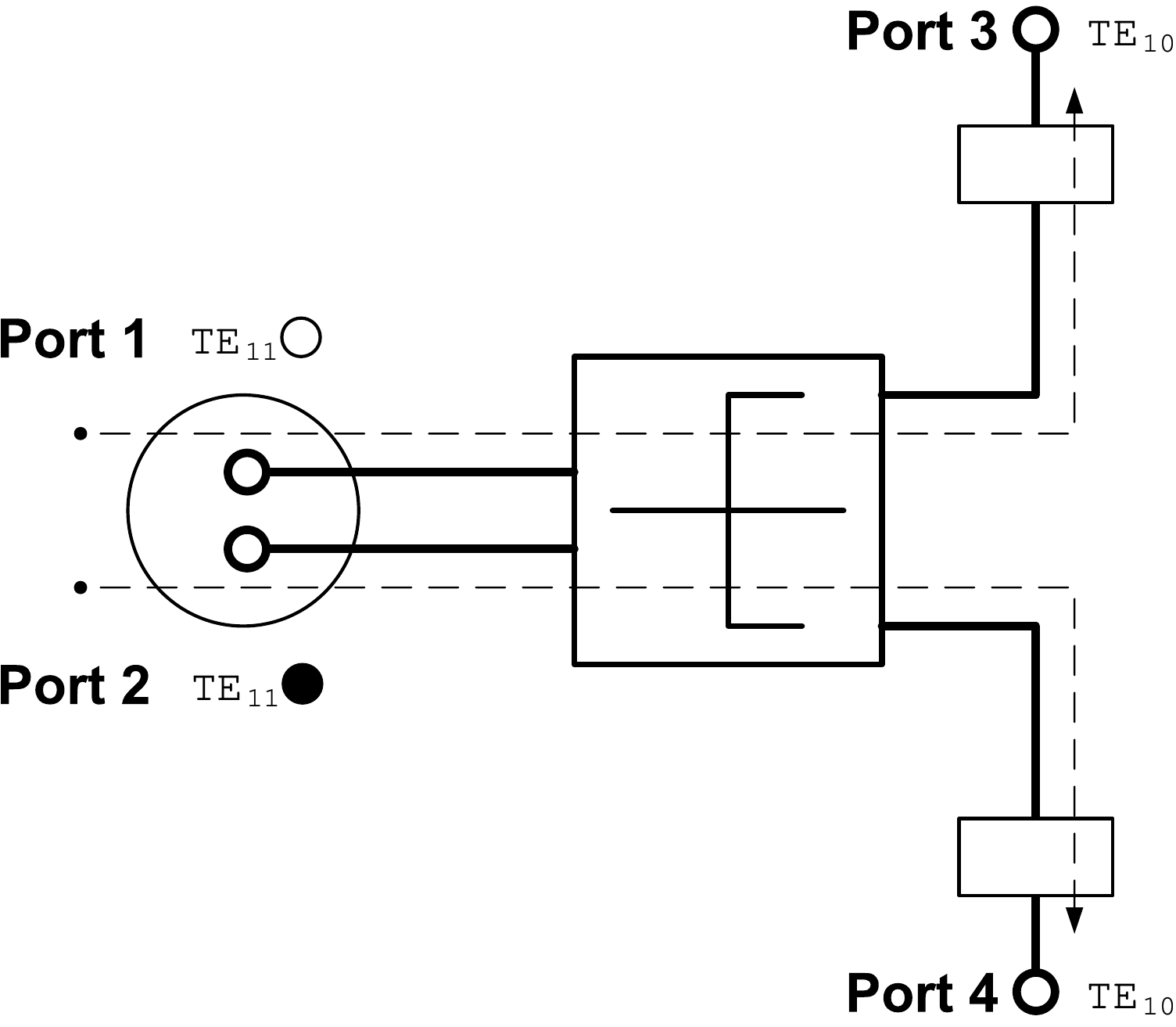}
 \caption{Electrical Scheme of the OMT adapted from Uher et al. \cite{Uher}. Port 1 and Port 2 are connected to
   the feed horn and are physically coincident even if distinct from the electromagnetic point of view.
   Port 3 and port 4 are the two outputs corresponding to orthogonal polarizations.}
\label{fig:omt_scheme}
\end{figure}

The OMT is a four port device (see e.g. Uher et. al., \cite{Uher}), where two ports share the same physical position, 
and correspond to two orthogonally polarized modes in circular waveguide,
as reported in figure 
\ref{fig:omt_scheme}. 
Port 1 and port 2 are in common and are connected to the feed horn, 
so that the signal coming from the telescope is split into two orthogonal components at 
port 3 and port 4, feeding two pseudo-correlation receivers. An ideal OMT would perfectly divide the components 1 
and 2 and transfer the signal to the ports 3 and 4 respectively, without cross-talk between components (perfect 
isolation and absence of cross-polarization), without reflections (perfect match) and without losses (perfect conductivity). 

For a propagating waveguide mode, where voltage and current cannot be uniquely defined, 
a generalized complex amplitude $a$ can be defined so that $P=a \cdot a^{*}$.
Calling $a_{i}$ the generalized amplitude of the signal at frequency $\nu$ entering the device from Port $i$, and 
$b_{i}$ that of the signal leaving the device from the same port, one can 
write the outputs as a function of the inputs with the matrix equation: 
\begin{equation}
b ={\bf S}\cdot a
\end{equation}
where $S$ is the Scattering Matrix. For the OMT the scattering matrix is a $4\times4$:

\begin{equation}
\bordermatrix{& \cr
& b_{1}\cr
& b_{2}\cr
& b_{3}\cr
& b_{4}\cr
}
=
\bordermatrix{& & & &\cr
& S_{11} & S_{12} & S_{13} & S_{14} \cr       
& S_{21} & S_{22} & S_{23} & S_{24} \cr
& S_{31} & S_{32} & S_{33} & S_{34} \cr
& S_{41} & S_{42} & S_{43} & S_{44}\cr
}
\cdot
\bordermatrix{& \cr
& a_{1}\cr
& a_{2}\cr
& a_{3}\cr
& a_{4}\cr
}
\end{equation}

It is readily seen that $S_{nn}$ is the reflection coefficient at port n, thus for example $\left|S_{11}\right|^2$ is
the fraction of power lost by impedance mismatch at port $1$, and  its reciprocal $RL_{(1)}=\left|S_{11}\right|^{-2}$ is
the return loss at port 1.  Similarly, $S_{31}$ is the transmission coefficient, and
$\left|S_{31}\right|^2$ the fraction of power transmitted, including impedance mismatch, 
ohmic losses and mode conversion along the path from port $1$ to $3$, whereas its reciprocal
$IL_{{31}}=\left|S_{31}\right|^{-2}$ is the insertion loss for port 3. 
The quantity $\left|S_{41}\right|^2$ represents the fraction of power transmitted from port 1 to 4, and its reciprocal  $XP_{41}=\left|S_{41}\right|^{-2}$ is the cross-polarization at port 4.
$\left|S_{34}\right|^2$ represent the fraction of power that can be transmitted across the output ports, and its reciprocal $IS_{34}=\left|S_{34}\right|^{-2}$ is the output isolation. 
$\left|S_{12}\right|^2$ represents the cross-polarized reflection (i.e. the fraction of power converted in the wrong polarization at the input port and reflected towards the horn).  Its reciprocal $IS_{12}=\left|S_{12}\right|^-2$ is the input isolation.
An ideal OMT, perfectly matched, with no cross polarization and insertion losses, exhibits the following 
Scattering Matrix: 
\begin{equation} 
S_{ideal}^{OMT}=
\bordermatrix{& & & & \cr         
  & 0  & 0 & e^{i\phi_1}& 0 \cr          
  & 0  & 0 & 0 & e^{i\phi_2} \cr          
  & e^{i\phi_1}  & 0 & 0 & 0 \cr          
  & 0  & e^{i\phi_2} & 0 & 0 \cr  }        
\end{equation}
being $\phi_1$ and $\phi_2$ the phase delays experienced by the wave traveling between port 1 and 3 and port 2 and 4 respectively. 
The matrix is symmetric because the device is reciprocal.

A similar notation is introduced here to represent the electromagnetic design requirements of LFI OMTs. Instead 
of using the scattering matrix $S$, we define 
\begin{equation}
T = 
\bordermatrix{& & & & \cr         
  & \left|S_{11}\right|^2  & \left|S_{12}\right|^2 & \left|S_{13}\right|^2 & \left|S_{14}\right|^2 \cr          
  & \left|S_{21}\right|^2  & \left|S_{22}\right|^2 & \left|S_{23}\right|^2 & \left|S_{24}\right|^2 \cr          
  & \left|S_{31}\right|^2  & \left|S_{32}\right|^2 & \left|S_{33}\right|^2 & \left|S_{34}\right|^2 \cr          
  & \left|S_{41}\right|^2  & \left|S_{42}\right|^2 & \left|S_{43}\right|^2 & \left|S_{44}\right|^2 \cr}          
\end{equation}
so that, denoting with $RL_{(i)}$ the return loss at port $(i)$, with $IL_{{ij}}$, $XP_{ij}$, and $IS_{ij}$ respectively 
the insertion loss, the cross-polarization, and the isolation between port $(i)$ and port $(j)$, we obtain: 

\begin{equation}
T = 
\bordermatrix{& & & & \cr         
  & 1/RL_1  & 1/IS_{12} & 1/IL_{13} & 1/XP_{14} \cr          
  & 1/IS_{21} & 1/RL_2 & 1/XP_{23} & 1/IL_{24} \cr          
  & 1/IL_{31}  & 1/XP_{32} & 1/RL_3 & 1/IS_{34} \cr          
  & 1/XP_{41}  & 1/IL_{42} & 1/IS_{43} & 1/RL_4 \cr}          
\end{equation}

The Electromagnetic requirements have been setup using these quantities and are reported in Table \ref{tab:em_requirements}.
  
\begin{table}
\caption{Electromagnetic requirements of the LFI's OMT.} 
\label{tab:em_requirements}
\centering
\begin{tabular}{l c c c} 
\hline\hline
                     
     \hspace{2cm}          & 30~GHz   & 44~GHz & 70~GHz \\
\hline            
Bandwidth (GHz) \dotfill  & 27--33 & 39.6--48.4 & 63--77 \\
IL @ 20~K\dotfill &  $< 0.15$~dB          &   $< 0.15$~dB  & $< 0.15$~dB \\
IL @ 300~K\dotfill &  $< 0.30$~dB   &  $< 0.30$~dB  & $< 0.30$~dB   \\
RL\dotfill             &  $> 20$~dB       & $> 20$~dB & $> 20$~dB \\                  
XP\dotfill &      $> 25$~dB & $> 25$~dB & $> 25$~dB        \\
IS\dotfill & $> 40$~dB & $> 40$~dB       & $> 40$~dB        \\       
\hline                                   
\end{tabular}
\end{table}

The Insertion Loss directly impacts the overall radiometer sensitivity since the OMT acts as a microwave 
attenuator at a physical temperature $T_0$ (in flight condition $T_0\sim 20$~K). The noise temperature of the OMT is then: 
\begin{equation}
T_{N}^{omt} = (L_{omt} - R_{omt}/(R_{omt}-1))\cdot T_0
\label{eq:noise}
\end{equation}
where
\begin{equation}
L_{omt} =  \left [ 10^{IL_{dB}/10}\right ]
\label{eq:loss}
\end{equation}
and 
\begin{equation}
R_{omt} =  \left [ 10^{RL_{dB}/10}\right ]
\label{eq:loss}
\end{equation}

$IL_{dB}$, $RL_{dB}$ are given in Table \ref{tab:em_requirements}.
The result is a $7\%$ increase ($\approx1.4~K$) in the radiometer noise temperature.
The requirements on return loss stem from the need of reducing as much as possible the ripple in the receiver passband, since the input return loss of the pseudo-correlation receivers
cannot be very high, and from the need of reducing backscattered power that could possibly reenter the system and increase the stray radiation. Obviously it is also necessary to effectively collect as much  as possible of the available power.

Isolation and cross-polarization have a direct impact on the instrument capability of detecting highly polarized spectral components. Although not specifically set for CMB polarization (they were set at the very early stage of the project), these requirements satisfy the scientific goal of the mission. 


\section{Design and Manufacturing}\label{fabrication}
At the very beginning of the project the symmetric OMT configuration (Boifot, \cite{boifot}; Wollack, \cite{wollack1};
Wollack et al., \cite{wollack2}) was studied as a possible solution thanks to its wide-band performance.
	 
This solution was abandoned for incompatibility with the mechanical requirements, particularly size and weight. An asymmetric design was then selected (Uher et al., \cite{Uher}).

This design is composed of a common polarization part (connected to the
feed horn) and main and side arms in which the two polarizations are separated.
The use of a septum to enlarge the bandwidth was considered and then discarded, due to the criticality of its positioning inside the structure and also because of the extremely small thickness required at the LFI frequencies.

Six main sections (a-f) have been identified and optimized during the design phase, each one corresponding to a well identified functionality.

\begin{itemize}
\item[a.] Circular to square waveguide transition. This section transforms the two TE$_{11}$ orthogonal circular
waveguide modes (one at Port 1 and the other at Port 2) into the two orthogonal modes TE$_{10}$ and TE$_{01}$
propagating in square waveguide.
\item[b.] Square waveguide section. Here the two principal TE$_{10}$ and TE$_{01}$ modes are well established,
ready to be divided into the two arms of the OMT.
\item[c.] Matching section between the TE$_{10}$ square waveguide mode and the main arm TE$_{10}$ fundamental
rectangular waveguide mode.
\item[d.] Matching section between the TE$_{01}$ square waveguide mode and the side arm TE$_{10}$ fundamental
rectangular waveguide mode.
\item[e.] Twist to correctly interface the OMT with the Front End Unit Module flange. A stepped geometry
was chosen to be compatible with the maximum allowable volume.
\item[f.] Stepped bend to adapt the OMT to the mechanical Front End Unit Module interface.
\end{itemize}

\begin{figure}
\centering
\includegraphics[width=6cm]{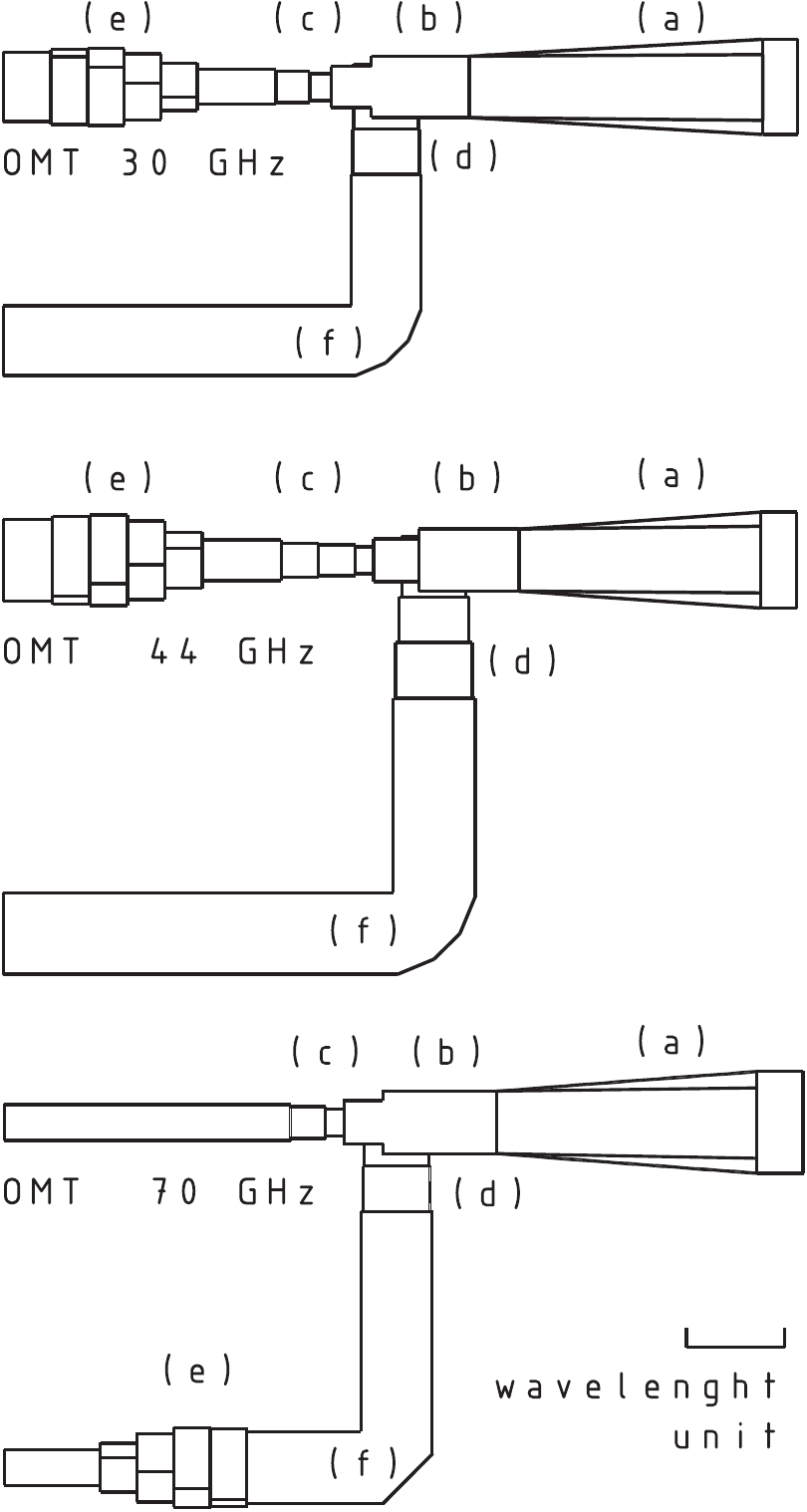}
   \caption{Sketch of the three OMTs with different parts identified by letters (see text for details). Top: 30 GHz OMT;
   Centre 44 GHz OMT. Bottom: 70 GHz OMT. The three drawings are normalized to centre wavelength, 10mm, 6.82mm,
   and 4.29mm respectively. The tree designs show that the normalized length is almost the same, even if not precisely,
   and that the 70 GHz OMT has the twist on the side arm due to the mechanical interface contraints.}
   \label{fig:omt_sketch}
\end{figure}

Figure \ref{fig:omt_sketch} shows these parts, normalized to the wavelength, in the three OMT types.
The three designs are not scaled, but they are similar. At 70 GHz the only
difference is on the twist which is on the side arm instead of the main arm of the OMT.
 The twist is a critical part of the OMT design, because it has very small size, i.e. 11 mm in the 30 GHz case, for a torsion of 90$^{\circ}$ and the orientation of the receiver ports require that it be on the main arm at 30 and 44 GHz, which makes matching more difficult.
In order to meet the goal of 20 dB return loss over the 20\% bandwidth, four possible configurations have been considered for the design of the twist, varying the number of the steps (one to four). At the end of the analysis, the configuration with four steps was chosen, since it grants a return loss close to the requirements in the operational bandwidth.

The design was frozen after several iterations of simulation and optimization using commercial software.\footnote{CST Microwave Studio (www.cst.com) and HFSS (www.ansoft.com)}

\begin{figure*}
\centering
\includegraphics[width=0.75\textwidth]{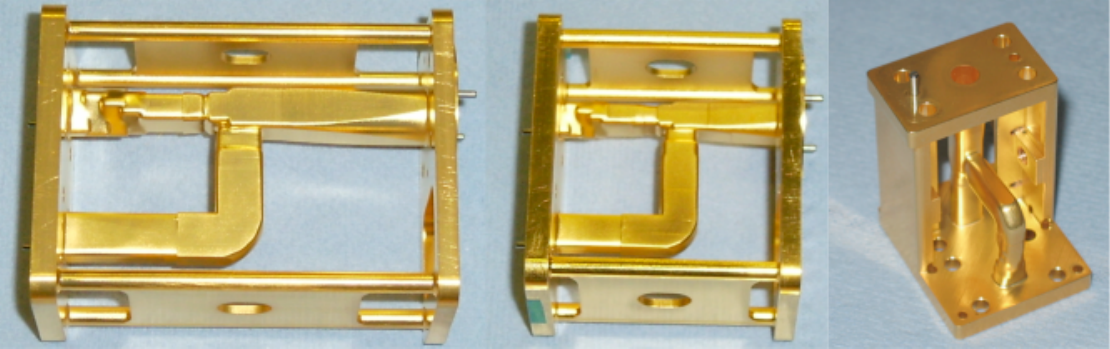}
   \caption{The three Ortho Mode Transducers. Left 30 GHz; centre 44 GHz; right 70 GHz.
   The three pictures are not in the same scale. }
   \label{fig:OMT3d}
\end{figure*}

The electromagnetic design was tested at first on a prototype realized in Al6061 with direct machining of the two separate half-shells. They were coupled together with screws and dowel pins in order to allow precise mounting. Electromagnetic performances were found to be critically dependent on the assembly pressure. Moreover, the accuracy in the realization of the twist steps was insufficient.
 The full set of OMTs was then built with an electroforming technique; a pure aluminum mandrel is directly machined in order
to obtain the inverse pattern of the final object. This master is then placed in a low current bath, where it collects
the copper ions. The final dimensional tolerances are thus those of the master, since the copper ions reproduce the
master profile at molecular level. The removal of the master is eventually done with a chemical corrosive solution
that melts it. Finally the OMTs are gold plated (figure \ref{fig:OMT3d}).

\begin{figure}[!h]
\begin{center}
\includegraphics[width=0.6\textwidth]{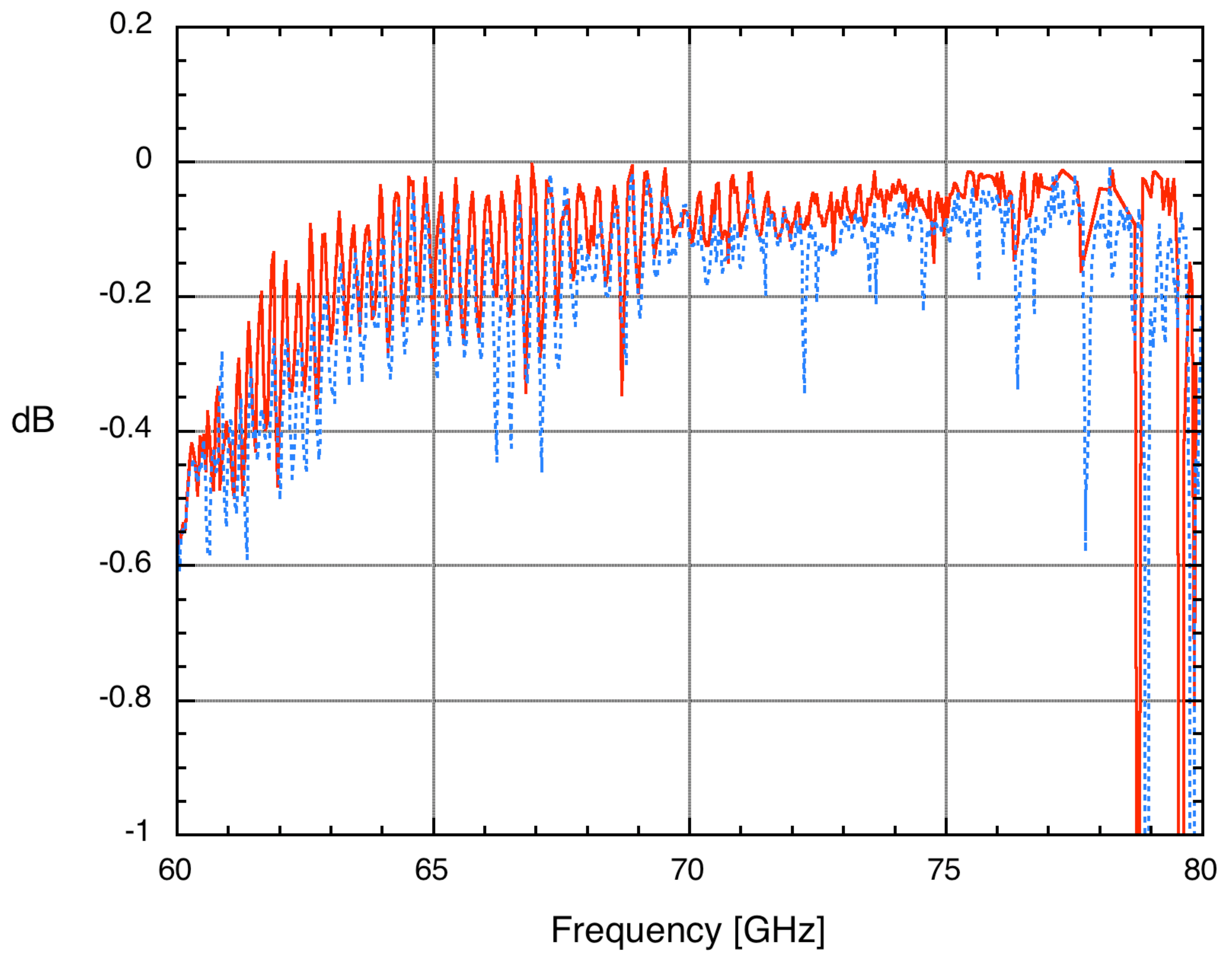}
\caption{Transmission coefficient of two Qualification Model OMT, main arm: one is internally gold plated (blue dashed line) while the other is not (continuous red line).}
\label{fig:IL_gold_nogold_QM}
\end{center}
\end{figure}

The OMT design takes into account the mechanical tolerances by checking the sensitivity of the simulated scattering parameters to variations in critical dimensions within the tolerance declared by the manufacturer. Furthermore, the simulations were repeated on the basis of the measured dimensions of the mandrels, considering that the electroformed pieces are their inverse replica.\newline
Four Qualification models OMT were built (1 at 30 GHz, 1 at 44 GHz and 2 at 70 GHz), obviously all eleven Flight Model OMT (two at 30, three at 44 and six at 70 GHz) and also 3 Flight Spare OMT (one for each frequency band) ready for use in case of failures of flight components. During the QM phase, only one of the two 70 GHz OMTs was internally gold plated, in order to compare the performance of the two components and to ensure that the deposition does not significantly alter the electromagnetic properties, as shown in figure \ref{fig:IL_gold_nogold_QM}.

\section{Qualification Campaign}\label{tests}
The qualification campaign included RF measurements on OMT stand alone and the vibration tests. In the early 
stages of the project RF tests were foreseen at ambient and cryogenic (20 K) temperature. For scheduling reasons 
the RF cryogenic test were not addressed. 
Since the start of tests, the scattering parameters of the OMTs were measured, namely the 
transmission and reflection coefficient of both arms but also the isolation between the two arms in order to assess the extent of separation of the two related radiometers of the same Radiometric Chain Assembly (RCA).
All tests have been performed at room temperature; the estimation of performances at cryogenic temperature will be discussed with the test results. The design of the OMT was optimized trying to obtain the best performance in the LFI  
frequency band (in fact, outside that band, OMT's performances degrade quickly). The extreme difficulty in meeting the requirements over the full bandwidth for the RL, while complying with the severe size constraints, became clear since the first tests on the prototype, named Elegant Bread Board (EBB).

The main constraints encountered in trying to meet requirements  were imposed by the small space available on the focal plane unit for the OMT. The final design guarantees important improvements in reflections performance with respect to the EBB prototypes, even if the requirement is not always met, as will be shown in section \ref{results}.

\subsection{OMT-FH integration: vibration tests}

After the complete electromagnetic characterization, each OMT was integrated with the corresponding FH. The assembly was then vibrated at 
space qualification level: the experimental set up is described in  Guzzi, \cite{Guzzi}.
Vibration tests were performed with a computer-controlled shaker generating the motion with the requested characteristic. The  accelerometers, placed on the OMT and FH assembly, were used for feedback. Their recorded signals were amplified, digitized and stored for processing.
The test started with a search for resonances in the range 5--2000~Hz, once the assembly was firmly mounted on the shaker.
Compliance with the following conditions was requested for the vibration test to be passed: after each random vibration on every axis, the OMT should not show visible degradation and all the screws holding the OMT to the FH should be tight to the specified torque. Moreover, the mechanical resonance frequencies before and after vibration should be shifted by less than ±5\%, while the variation in acceleration should be smaller than ±10\%. 
Initially, it was decided that the measured reflection coefficient 
of the assembly should be taken as the electromagnetic parameter to check before and after vibration, because of its sensitivity 
to mechanical imperfections, in order to guarantee that the test was passed successfully. In this case, the OMT reflections are 
dominant with respect to those of the FH, so one could expect only small differences even under worst case. The Fourier Transform (FT) of the signal (i.e. the time domain sequence of reflection peaks) was used as a tool for the comparison, since it allows to pinpoint the spatial location inside components of any differences between pre-- and post--vibration reflections. In the very few cases where differences were seen, they were found in the coupling between OMTs and flange adapters due to a small inaccuracy  in the pre--vibration measurement setup. It was discovered later that a more sensitive
test could be made, measuring also the radiation pattern of this assembly; in fact, especially at higher frequencies, the far-sidelobe pattern of the 
FH+OMT shows extreme sensitivity to the relative alignment between components; 
thus it can be used as additional information to reveal any possible variation that may have occurred in the connection between the FH and the OMT.   

All 11 assemblies FH+OMT were vibrated: the first 70~GHz assemblies did not pass the tests, since after vibration 
they showed visible cracks around the flange connecting them with the FH. This problem was encountered 
only at 70~GHz since, even though the physical dimensions are smallest, the FHs are actually the heaviest, being made of gold-plated copper instead 
of aluminum. After this problem was found, the broken components were replaced and all the 70~GHz FH and OMT 
flanges were reinforced. For this reason,
it was necessary to characterize once again all the reinforced components. At the end of these steps, all the 
assemblies were vibrated again successfully. 

\subsection{Performed tests and measurements technique}\label{Performed tests}

The test matrix was defined at the beginning of the program phase and found adequate for all the development stages.
The experimental set-up was progressively improved to reach good repeatability and an overall error bar lower than 0.1 dB and few degrees (amplitude and phase) in measurements  of (copolar) transmission coefficients.
Several auxiliary components were necessary to couple the OMT with the measurement set-up. Special flange adapters,  circular to rectangular waveguide transitions (see ports 1 and 2 in Fig. \ref{fig:omt_scheme}) and twists were tested separately before use.

All tests were performed
using a Vector Network Analyzer (VNA)\footnote{AB millimetre, http://www.abmillimetre.com} in a 4S arrangement, with two source and two receiver ports, coupled to the device under test with two directional couplers.

\begin{figure}[!h]
 \begin{center}
\includegraphics[width=0.6\textwidth]{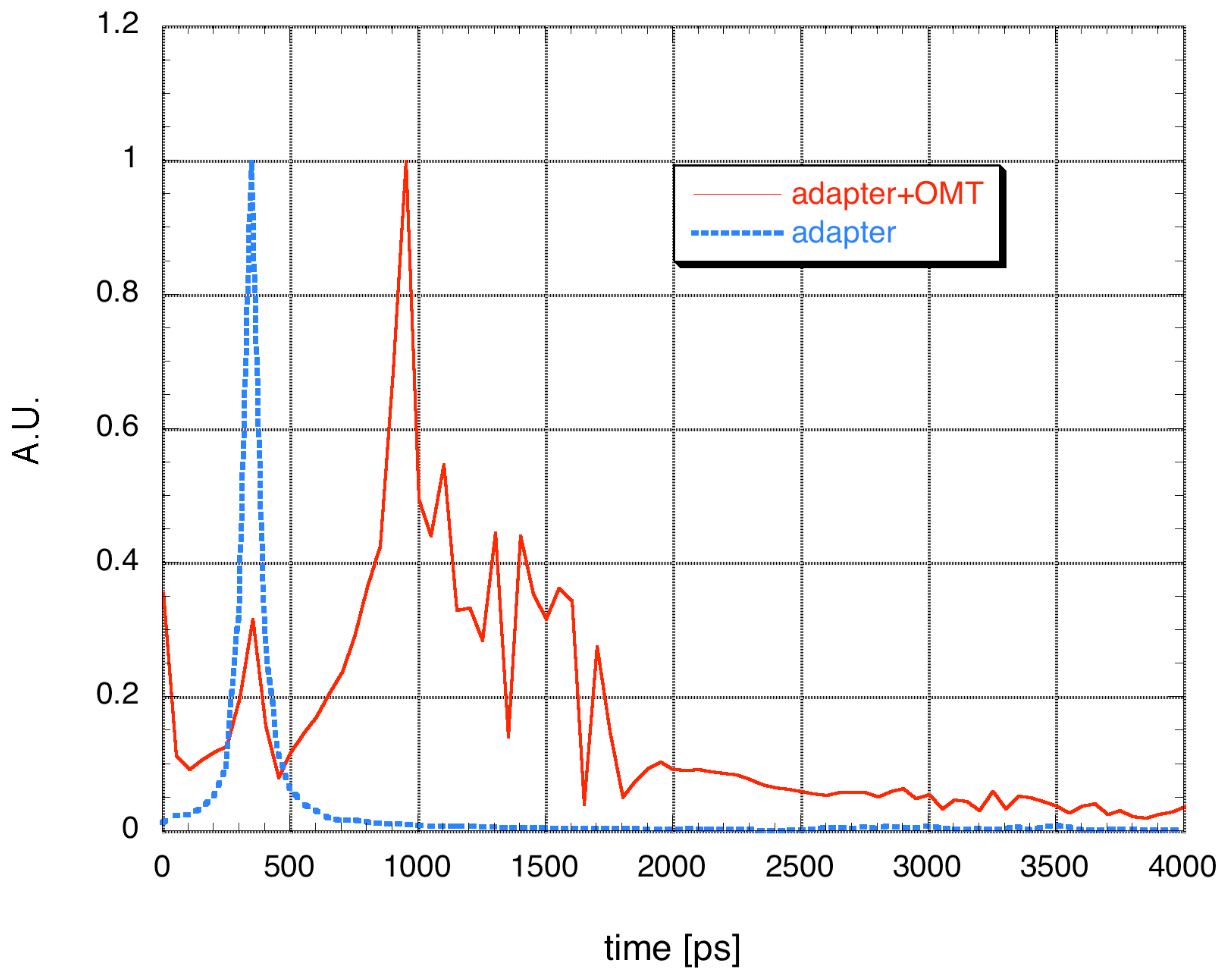}
\caption{\emph{FT of the frequency sweep (normalized to maximum) performed when one 70GHz adapter is terminated with a short (dashed line) and when the same adapter is connected to the OMT (continuous line): it is possible to distinguish the reflection peak due to the adapter and to remove it from OMT data.}} \label{fig:FFT}
 \end{center}
 \end{figure}

The Scattering parameters (S-parameters) were measured in a two-port configuration, 
using two directional couplers (and a twist when necessary). Full 12-term calibration with Thru--Short--Variable 
Short--Variable Load (see e.g. Engen, \cite{Engen}) was used in this case.
The unused arm of the OMT was generally terminated with a matched load, even if this
had no impact on results, given the excellent 
isolation between the two arms. 
In order to guarantee mechanical stability, each directional coupler was mounted on a suitable support plate that could be precisely positioned with (manual) linear stages. 
Isolators were used
on all the VNA heads, and 6~dB attenuators to further improve the output match of the generator heads. High phase stability cables\footnote{GORE$^{TM}$ VNA Microwave / RF Test Assemblies} were used between the VNA and the two heads that had to be moved to insert the device under test. With this experimental configuration it was possible to measure the phase and amplitude of transmission and reflection coefficients for both arms of the OMT.
The same arrangement was used for testing (pairs of) flange adapters. In this way it was possible to obtain a more precise evaluation of the losses of the adapters, while they have been also tested in a one port configuration as single pieces, terminating them with a short in order to evaluate their 
electrical length (i.e. propagation delay). 
Since the adapters have by definition non--standard flanges on one side, it was not possible to include 
them in the instrument calibration, and their contribution had to be subtracted 
from the results, assuming perfect identity between the members of a pair. 
Also standard twists were tested in pairs, since the 90 degrees rotation of the instrument heads required to accommodate a single twist might have been more detrimental to measurement precision than the effect of the twist.
The reflection coefficient of the Feed Horn--OMT assembly was measured in a one port  configuration after integration of the two units. 
One port calibration with Short--Variable Load--Fixed Short was used in this case as a standard.
Isolation measurements were made on the assembly FH+OMT (with the FH pointing at an absorber) 
using a simple transmit-receive arrangement and only a correction of the response of the VNA. 
This was justified, since the measured quantity was so low that the small additional correction due to full calibration was negligible.
Transmission coefficient results were corrected for the presence of adapters simply using their average loss with the measured phase lag. 
Reflection coefficients were corrected using
time domain filtering (time-gating): the FT of a frequency sweep gives a sequence of peaks as a function of time, representing reflections from
different parts of the device under test (see figure \ref{fig:FFT}). 
Filtering (i.e. zeroing out) the unwanted contribution of adapters and additional components and transforming back to 
frequency allows correction of the data. The main difficulty in the process is that the waveguide is dispersive and time (i.e. space) resolution is limited by the small extent of the
frequency sweep. Both effects make the peaks wider  (about 5--8~mm equivalent resolution, depending on the frequency band). To have a reasonably precise identification of the location of the interface of all components,  
the electrical lengths of the OMTs and all other components and adapters used during the tests were measured in the one port configuration described above, 
terminating the device under test with a reflective Short, that allowed the sharpest possible identification of all port locations.  
Multiple reflections did not represent a significant problem for adapters on the generator port, because their return loss was higher than that of the OMTs, 
and peaks of multiple reflections between adapter, VNA, adapter were very small with respect to those of the OMTs. The effect of the adapter on the detector port was in principle more difficult to remove, because of any multiple reflections inside components and between them and generator--port adapters. Visual inspection of time--domain echoes (as in fig.  \ref{fig:FFT}) does not show significant evidence of such effects. One--port measurements (with one adapter only) were always made on the OMT+FH assemblies, and they are in general agreement with the two--port ones, supporting these considerations.
\section {Results}\label{results}
An overview of the results of the electromagnetic tests is described here for each type of measurement: data were acquired over a frequency range as large as possible and, where available, results are compared with simulations. The transmission 
coefficients
are used to evaluate departures of the OMT from ideal performances, in order to establish the polarization capability of LFI (Leahy et al., \cite{Leahy2009}).

\subsection{Transmission Coefficient: co-polar term}

\begin{figure*}
\centering
\includegraphics[width=0.8\textwidth]{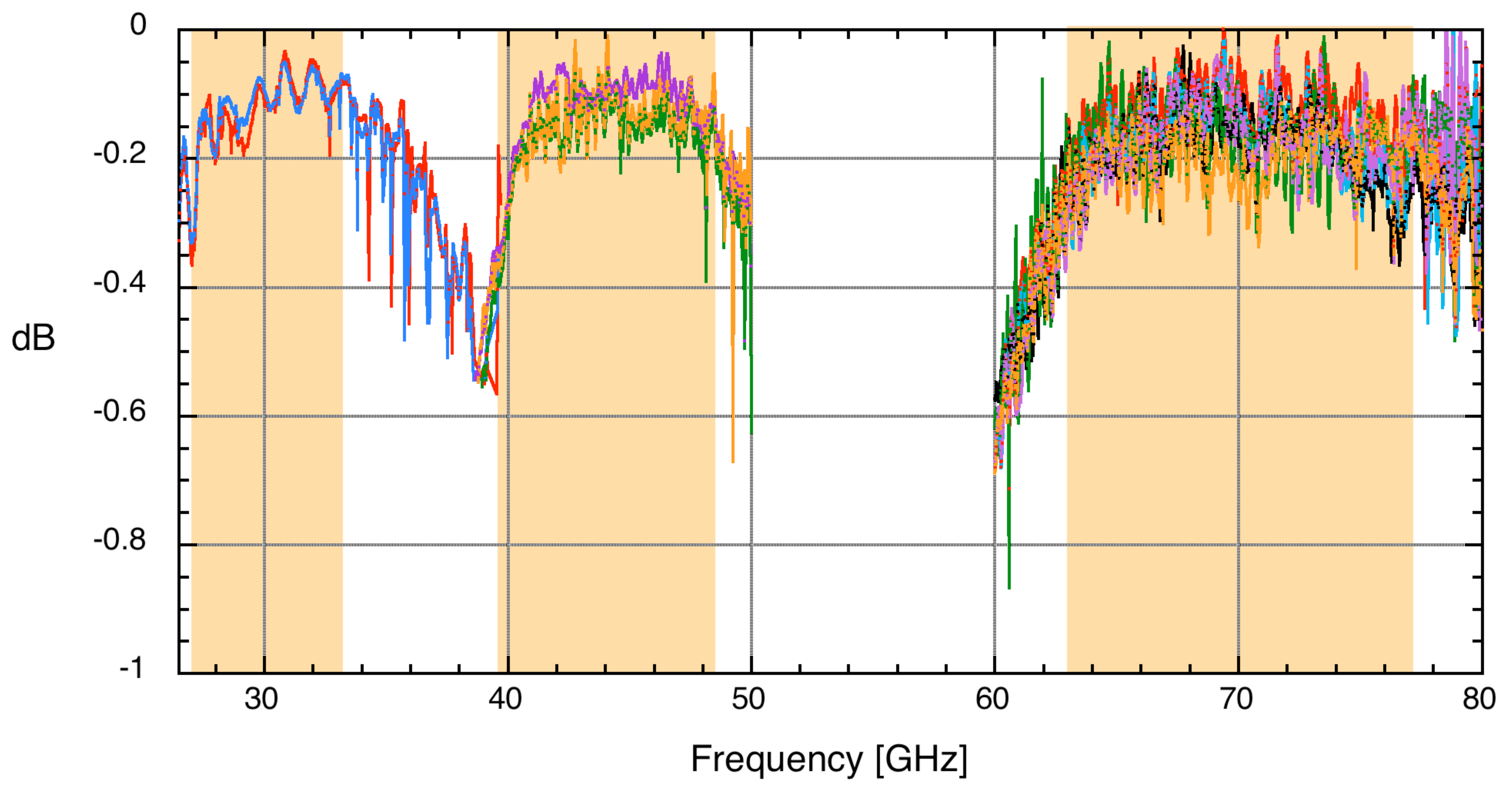}
   \caption{Amplitude of transmission coefficient (Main arm, $\left|S_{31}\right|^{2}$) 
of all the eleven Ortho mode transducers: the LFI frequency bands are evidenced in the figure.}
   \label{fig:S21main}
\end{figure*}

\begin{figure*}
\centering
\includegraphics[width=0.8\textwidth]{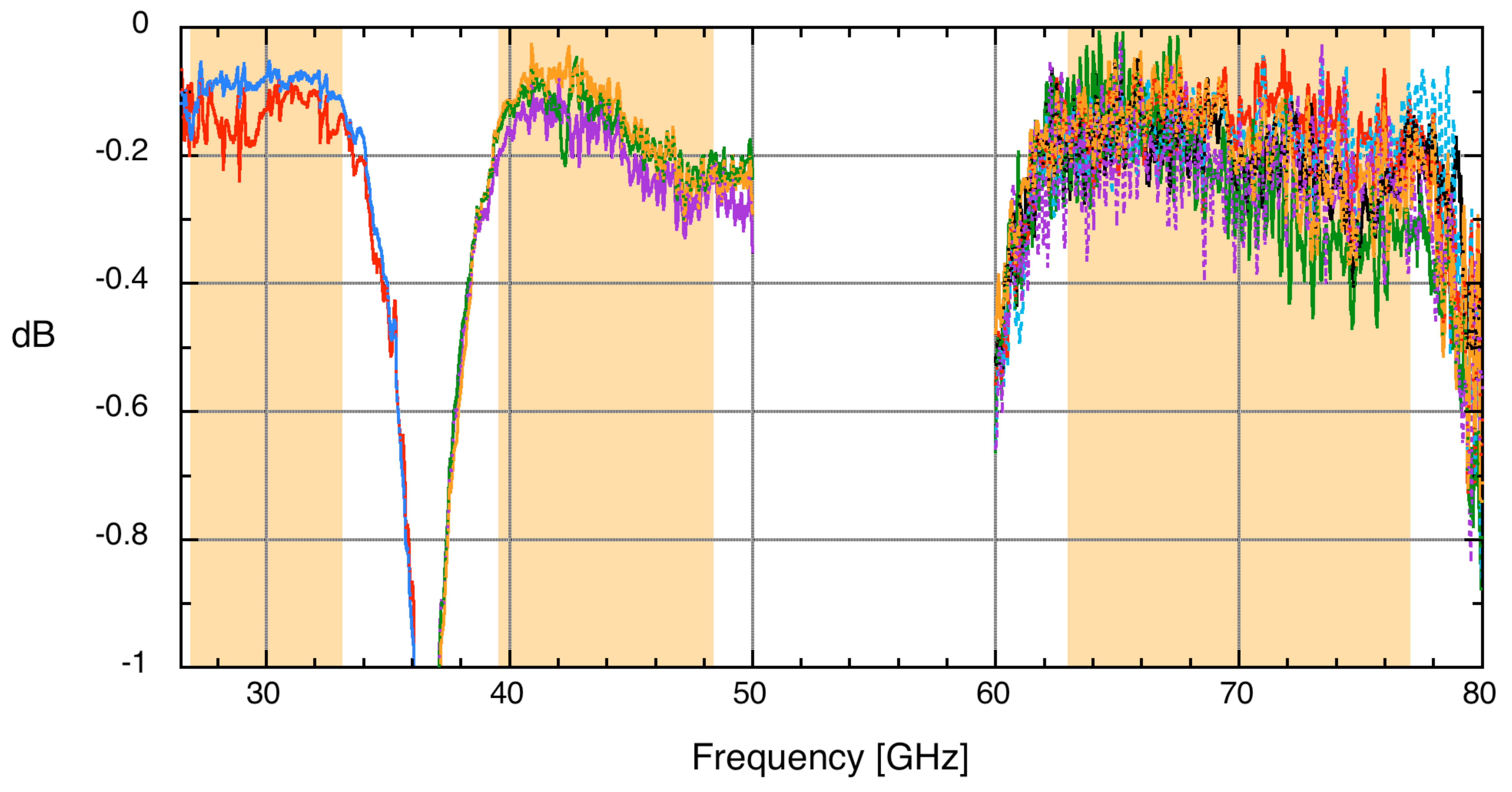}
   \caption{Amplitude of transmission coefficient (Side arm, $\left|S_{42}\right|^{2}$) 
of all the eleven Ortho mode transducers : the LFI frequency bands are evidenced in the figure.}
   \label{fig:S21side}
\end{figure*}

The insertion loss of the OMTs was determined subtracting the contribution of all the components necessary for the tests. It was always measured inserting the signal both from the circular and the rectangular port. 
Since the OMTs are reciprocal components, only data obtained for one direction are shown 
here, in order to simplify the plots (generator port on the rectangular waveguide), even if both $S_{mn}$ and $S_{nm}$ were always measured.
At 30 GHz, since the OMT has standard flanges, it was only necessary to 
remove the contribution due to the circular to rectangular transition, and also the twist for the main arm. The measured amplitude of the transmission coefficients
for the main/side arm over the 3 LFI frequency bands are reported in figure \ref{fig:S21main} and  \ref{fig:S21side} (amplitude).
All measurements were performed over larger frequency bands (26--40, 33--50 and 60--80~GHz) than those of LFI: the goal was to perform all tests over the widest possible bandwidth, in order to acquire as 
much information as possible for every component as well as to obtain the best spatial resolution for the 
FT.
As a first consideration, the similar behaviour of the OMTs is quite apparent in all frequency bands. They 
all exhibit small losses, largely meeting the room temperature requirement at 30 and 44~GHz. It is also interesting to note that
losses are in line with cryogenic requirement even at room temperature, especially in the 30 and 44~GHz channels. 
Thus, even if there are no data available in cryogenic conditions, the room temperature measurements give confidence for proper operation at 20~K. 
In general it was observed that insertion loss increases significantly outside the specified band. In particular at 33~GHz (approaching the cut off for the 44~GHz channel) 
it reaches about $35$~dB, because of the large reflections.
The LFI 70~GHz channel is the one with the greatest number of RCAs (6) and since it requires 6 OMTs it is the only one with a statistical spread. Nevertheless, even at 70~GHz, all OMTs show almost the same behavior, meeting the 
300~K requirements and being very close to the 20~K ones.
The widest possible band achievable was only 60--80~GHz in this case, less than the full waveguide band, because 
the available VNA heads are in WR15 (50-75~GHz) waveguide and their efficiency decreases quickly outside their intended frequency range. 
The insertion loss of one of the 70~GHz OMT, side arm, at room temperature is marginally outside the requirement at operation temperature over a part of the band; however it reaches 
a level of about 0.35~dB, only 0.05~dB outside of  requirement, a value comparable with the measurement error (less than 0.1~dB) and only over a 
partial fraction of the band. 

\subsection{Reflection Coefficient}

\begin{figure}[!h]
\begin{center}
\includegraphics[width=0.7\textwidth]{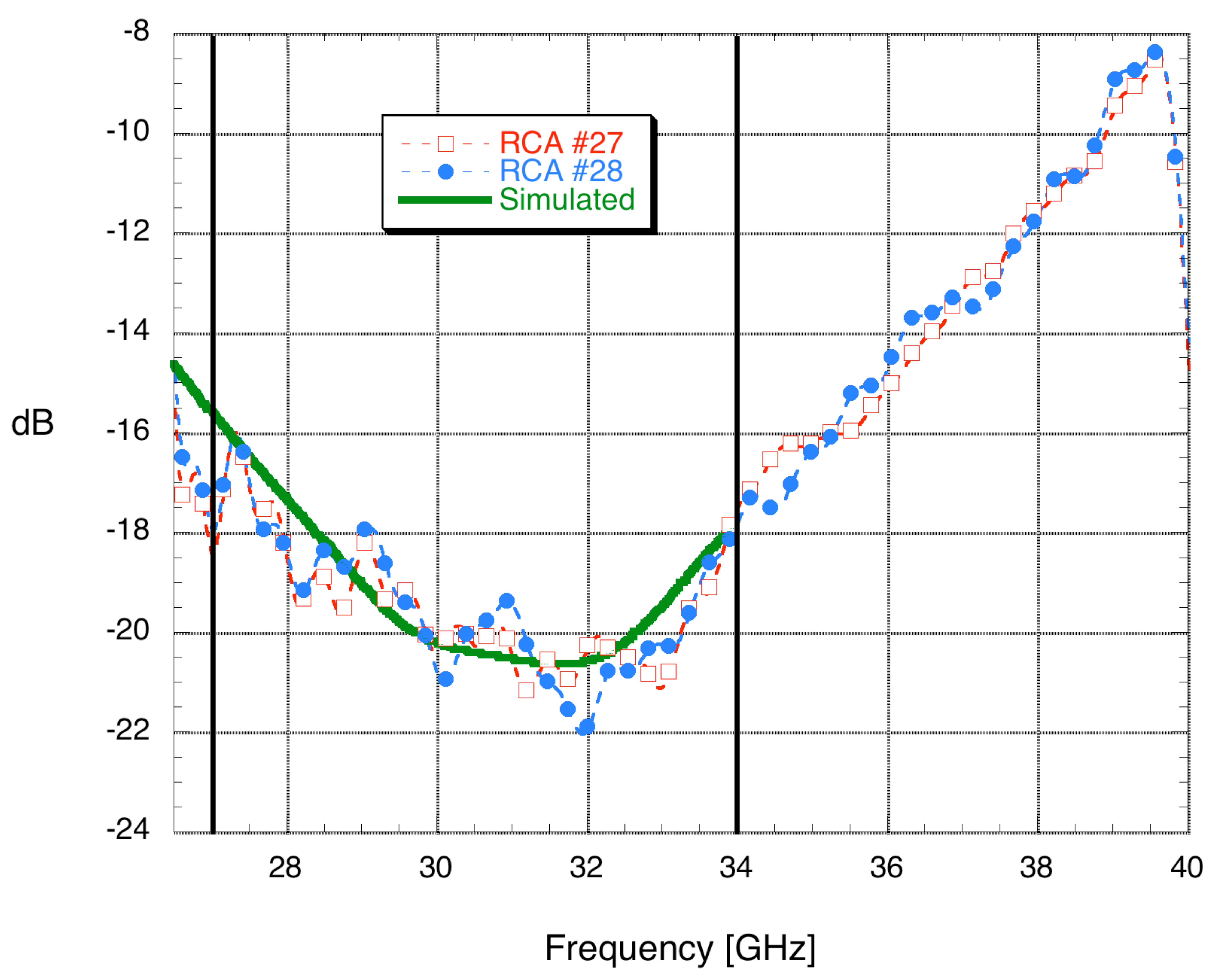}
\caption{ 
Amplitude of Main arm reflection coefficient
of the two FM OMT @ 30~GHz: the two vertical lines represent the 20\% bandwidth.}
\label{fig:RL30main}
\end{center}
\end{figure}

\begin{figure}
\begin{center}
\includegraphics[width=0.7\textwidth]{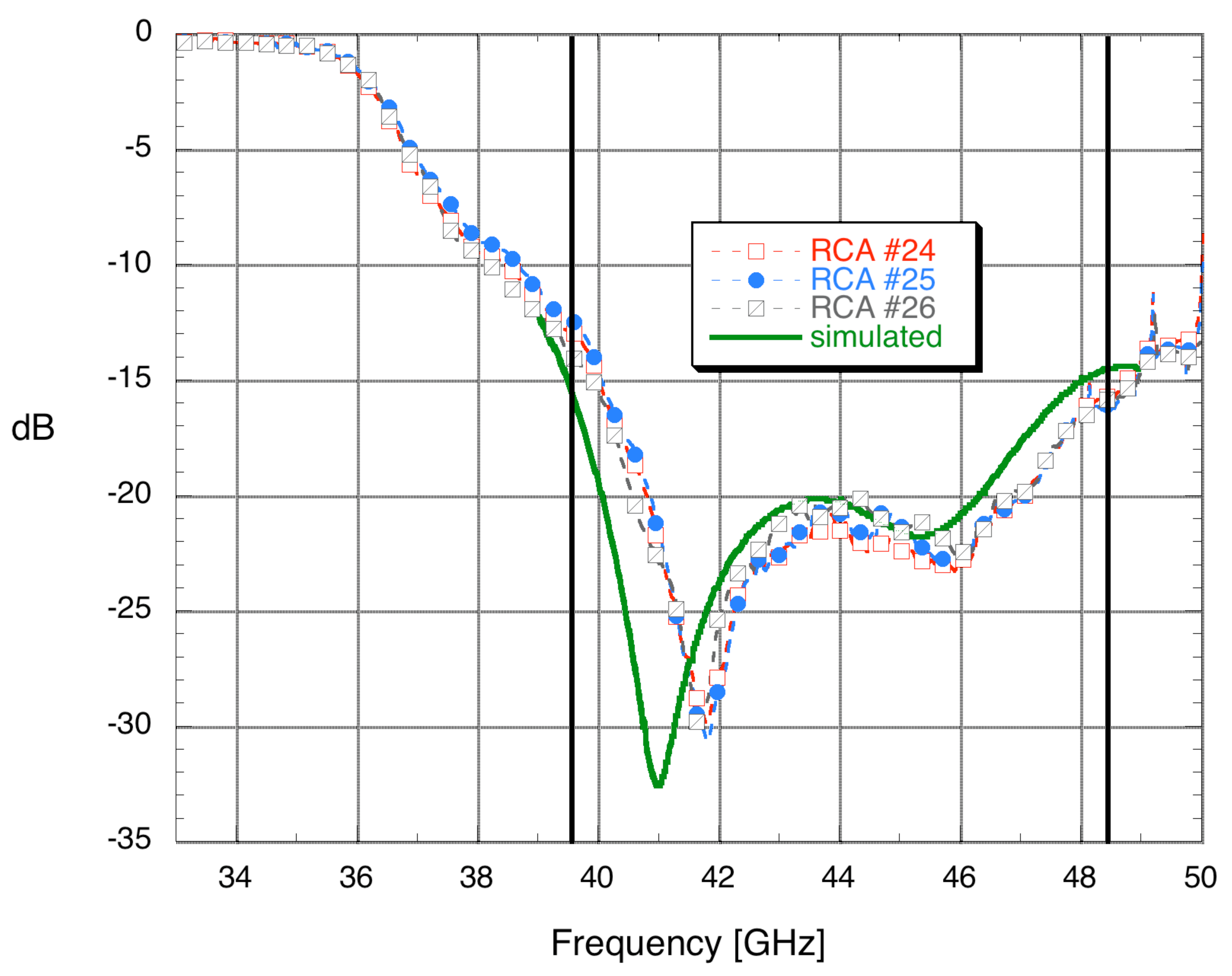}
\caption{
Amplitude of Main arm reflection coefficient
of the three FM OMT @ 44~GHz the two vertical lines represent the 20\% bandwidth.} 
\label{fig:RL44main}
\end{center}
\end{figure}

\begin{figure}
\begin{center}
\includegraphics[width=0.7\textwidth]{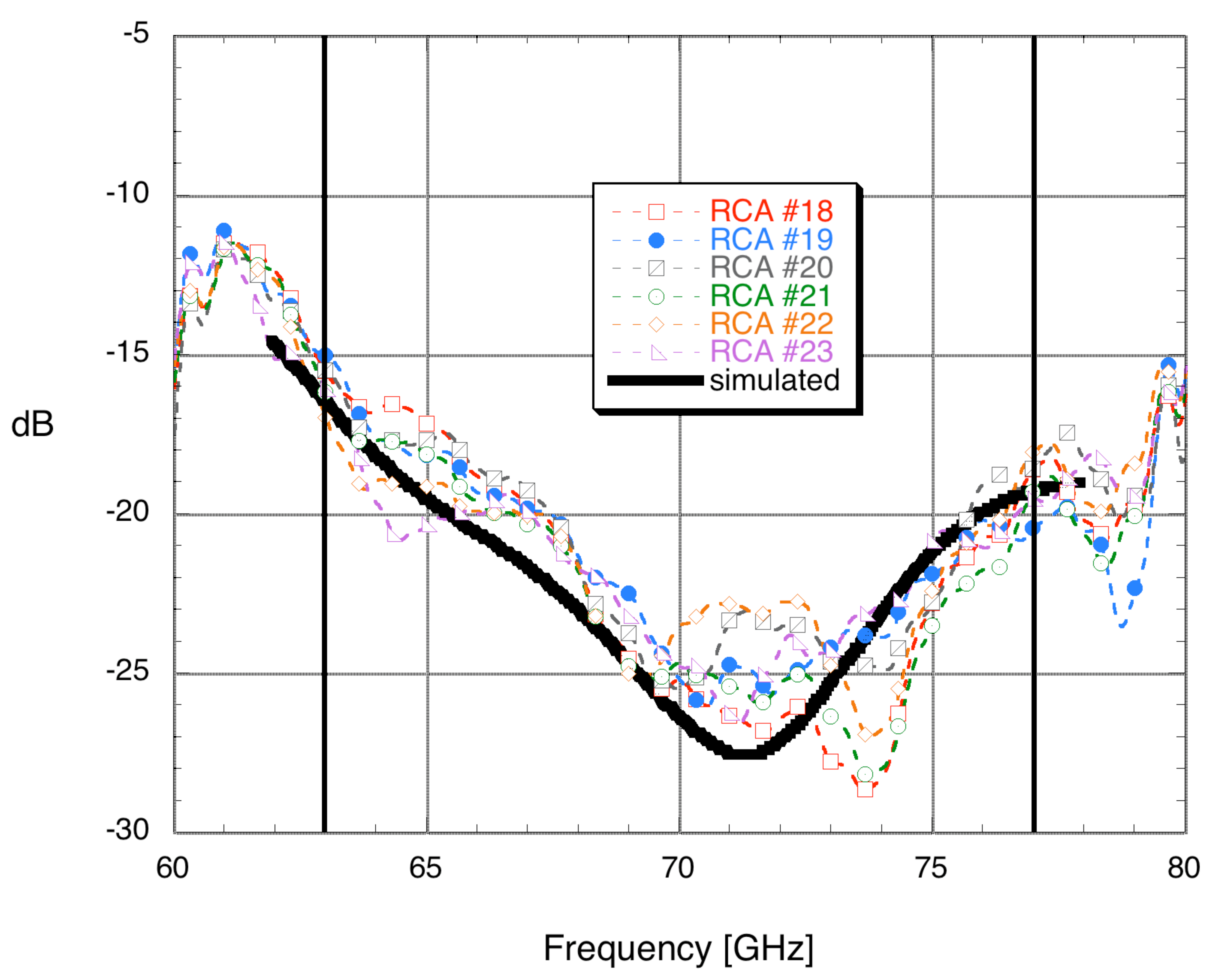}
\caption{
Amplitude of Main arm reflection coefficient
of the six FM OMT @ 70~GHz the two vertical lines represent the 20\% bandwidth.}
 \label{fig:RL70main}
\end{center}
\end{figure}

As previously mentioned, during the design phase the most problematic parameter to optimize was the reflection coefficient, 
mainly because of the mechanical and interface constraints. Even with the improvements adopted during the design 
process, it was not possible to reach the target specification of $-20$~dB over the full bandwidth. 
Simulations were below $-15$~dB everywhere for the design chosen, but the comparison between measurements 
and simulations was not satisfying after the first QM tests. A better agreement was found when the 
simulations were performed using the actual measured mandrel dimensions instead of the design specifications. 
The difference between the two simulations was up to 5~dB, which points to a critical tolerance in the design parameters, 
of course justified by the tight specifications. The quality control on the mandrels was therefore improved for the FM production. 
Anyway in figures \ref{fig:RL30main} to  \ref{fig:RL70main} only one representative simulation is shown for simplicity, even when they were made for every FM OMT. Moreover, only the main arm is shown here since normally (at 30 and 70~GHz) the side arm meets requirements, while the main arm reflection is the most critical parameter.
\newpage
The similarity in the measured reflection coefficient is very high for all the OMTs of the same band, and the agreement with simulation is usually good: as already said, the $-20$~dB requirement is always met on the 30 and 70~GHz side arm but not on the main arm, as predicted by calculations. 
The 70~GHz OMTs have probably the best performance in terms of reflection.
In fact, the amplitude of the reflection coefficient on the main arm of the 6 OMTs is always below $-15$~dB, and it is outside the specification only over a very limited 
portion of the band (figure \ref{fig:RL70main}). This is explained by the location of the stepped twist on the side arm, which  gave a lot more room for optimization. 
It is interesting to note that even at 70~GHz where there is more statistical spread, the measured parameters 
of all the 6 OMT are almost identical, which shows that the manufacturing process was under full control. 
\newpage
\subsection{Transmission Coefficient: cross-polar term}
\begin{figure}[!h]
\centering
\includegraphics[width=0.8\textwidth]{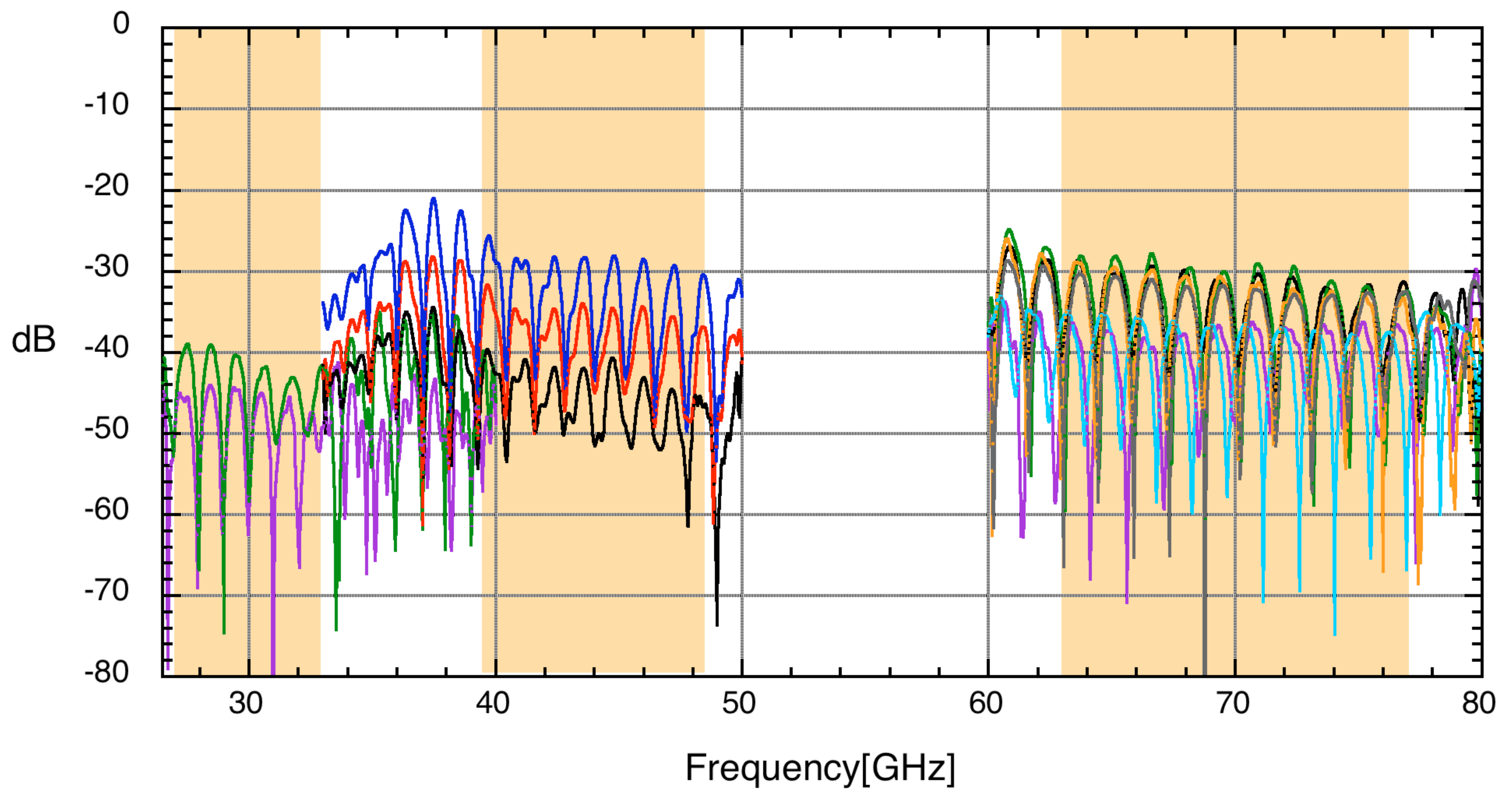}
   \caption{Side arm cross-polar response for all the OMTs (amplitude).}
   \label{fig:Xpol}
\end{figure}

The requirement on the amplitude of the cross polar term of the transmission coefficient was below $-25$~dB over the full operation band. Measurements were done exactly in the same way described for the co polar term, but in this case the circular port is fed with the wrong polarization for the arm under test.
As usual, the cross-polarization was measured inserting the signal both from the circular and the rectangular port, but given the reciprocity of the device, there is no difference (above noise) with the direction of propagation. 
Since both arms have a similar level of cross--polar response, only the Side arm data (generator on rectangular port) are shown in  figure \ref{fig:Xpol}.
\begin{figure}[!h]
\begin{center}
\includegraphics[width=0.7\textwidth]{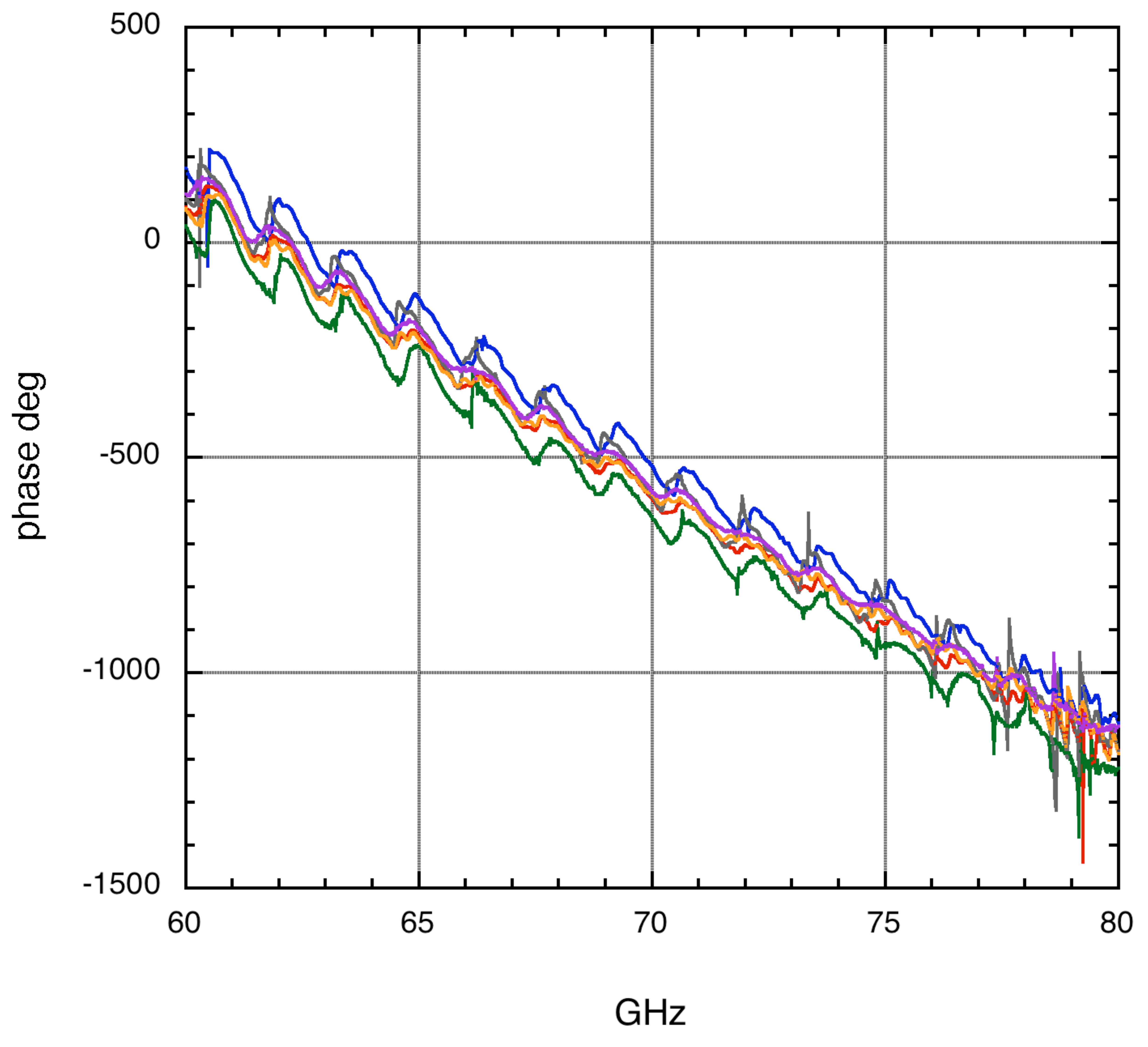}
\caption{Main arm cross-polar response for the six 70~GHz OMTs (phase). } 
\label{fig:Xpolphase}
\end{center}
\end{figure}
Both main and side arms of the two 30~GHz OMTs exhibit a cross-polar level well below the 
requirement, being the amplitude around $-40$~dB over the operational band. 
At 44~GHz, instead, the varying performance (by more than 10~dB) of these OMTs is quite apparent, even if they all comply with specifications (exactly the same situation happens for the main arm). There are a few differences also among the 70~GHz OMTs, but once again all of them meet the specifications. 
The phase of the cross--polar response is qualitatively very similar for all the OMTs. Figure \ref{fig:Xpolphase} shows the results for the 6 OMTs at 70~GHz (Main arm, generator on rectangular port).
The signal is noisy, but this is reasonable considering that the signal measured is very faint, at least at a level of $-30$~dB. These results show the difficulty of reproducing the 
same level of cross-polarization, but this is to be expected, 
since the measured levels are extremely low, and even very minor variations in the relative alignment of parts of the mandrel will cause significant changes in this small parameter. 
\subsection{Isolation}
\begin{figure*}
\centering
\includegraphics[width=0.8\textwidth]{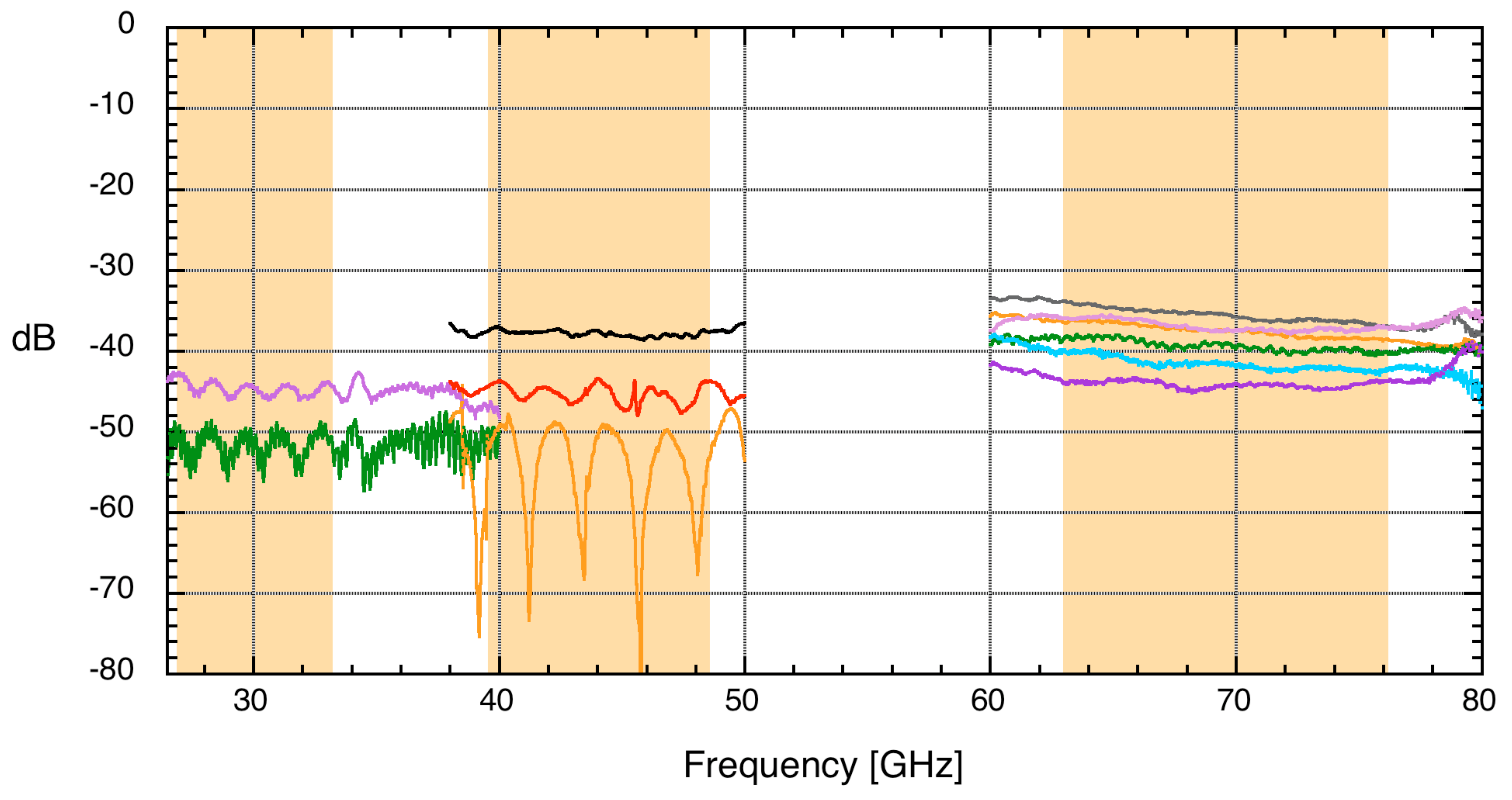}
   \caption{  $\left|S_{34}\right|^2$ (reciprocal of output isolation)
   of all the eleven Ortho mode transducers: the LFI frequency bands are evidenced in the figure.}
   \label{fig:iso_all}
\end{figure*}

The results obtained measuring the isolation between the main and side arms (port 3-4 figure \ref{fig:omt_scheme}) of all the OMTs are shown in this last subsection.
The measurements were made with the setup described in section \ref{Performed tests}. 
As in the other cases, both terms of the scattering matrix were measured, even if the device is reciprocal, as a verification of (random) measurement errors. The FH is facing an Eccosorb$^{TM}$ panel 
that absorbs the radiation. The results obtained
with the generator on the side arm are reported in figure \ref{fig:iso_all}.
The 30~GHz OMTs exhibit very good isolation, well below the requirement: the two 
units show a difference of about 4--6~dB almost constant over the frequency band. 
In the 44~GHz band, instead, there are clear 
differences among the three OMTs. First of all, the OMT of RCA 26 does not meet the requirement, 
even if the amplitude of the signal remains at a very low level, always below $-36/37$~dB. The other two units comply with specifications, even if at different levels (one around $-45$~dB and the other around $-50$~dB). 
The isolation of the 3 OMTs in this band is of the same order as the cross-polar insertion loss.
In the 70~GHz channel, only two out of the six OMTs meet the requirements. Nevertheless, they all show good isolation, the worst case being above $34$~dB 
inside the LFI frequency band. The spread in the measured values is between 7 and 10~dB.

\section{Estimated performance at flight conditions}\label{flight}

As shown in the previous section, the OMT's insertion loss is
extremely low. At 30 and 44~GHz, the OMT performances are already in line with the requirements at cryogenic temperature.  Nevertheless, a first rough estimation of the performance in flight conditions was extrapolated from room temperature data. The estimate was made treating the OMTs as rectangular waveguides, whose losses can be evaluated easily as a function of size, frequency and of course resistivity, which is the key parameter since it has the strongest temperature dependence. The resistivity of gold was used for the calculation, since each OMT is internally plated with a 2~$\mu$m layer of gold and skin depth is lower than 1~$\mu$m at all frequencies.  Using the resistivity of pure gold at room temperature underestimates losses, at the very least because of the finite surface roughness. An \emph{effective} resistivity was determined by fitting the measured data, including therefore the excess ohmic loss due both to surface conditions and to the difference in field structure with respect to a rectangular waveguide, but also including non-ohmic losses (reflections). Only in-band data were considered, since the performance of the OMTs quickly degrades outside the band of operation. 
Reflections are significant and can be the dominant loss mechanism: a $-15$~dB reflection accounts for nearly 0.15~dB insertion loss, and $-20$~dB for 0.05~dB.
Therefore the \emph{effective} resistivity is assumed to take the form of an \emph{additive} correction term $K$ to the bulk resistivity of pure gold. The correction term is assumed independent of temperature, which is an acceptable approximation for non-ohmic losses. 
The \emph{effective} resistivity is sometimes twice that of pure bulk gold, which would still fit the usual rule-of-thumb for the effect of surface quality, but the dominant loss term is not ohmic, as shown above. 
Since resistivity at 20K is less than two per cent of its value at room temperature, the estimated insertion loss for flight conditions is that of room temperature, multiplied by $\sqrt{\frac{K}{K+\rho}}$, i.e. in dB one should add 
$\frac{1}{2}[K_{dB}-(K+\rho)_{dB}]$, where $\rho$ is the bulk resistivity of pure gold.

Since ohmic losses are not dominant, the extrapolated performance at cryogenic temperature is usually very similar to that of room temperature. The correction is always less than 0.1~dB and somewhat larger for the side arm than for the main one. This is in qualitative agreement with its longer path, that enhances ohmic losses. In order to give a schematic view of the estimated losses, the mean value over the bandwidth has been calculated and compared with the mean value measured (table \ref{Tab:2}).

\begin{table}[h!]
       \caption{\emph{Mean value of the IL over LFI bandwidth estimated at 20K and measured at room temperature.}}
       \centering
       \begin{tabular}{lccccc}
       \hline\hline

  &	\multicolumn{2}{c}{IL @ T room}       & \multicolumn{2}{c}{IL @ 20~K} \\
&main [dB] &side [dB] &main [dB]  &	side [dB] \\
30~GHz\\
 RCA 27  & 0.11& 0.12& 0.10& 0.10 \\
 RCA 28 &0.12& 0.14& 0.11& 0.10\\
\\
44~GHz\\
RCA 24 &0.12& 0.19& 0.10& 0.17   \\
RCA 25 &0.16& 0.16& 0.15 & 0.14   \\
RCA 26 &0.14& 0.14& 0.13 &0.12   \\
\\
70~GHz\\
RCA 18 &0.17& 0.14& 0.16&  0.07   \\
RCA 19 &0.18& 0.20& 0.16 & 0.10   \\
RCA 20 &0.18& 0.24& 0.16 &0.16   \\
RCA 21 &0.16& 0.14& 0.15 &0.06   \\
RCA 22 &0.12& 0.18& 0.09& 0.09   \\
RCA 23 &0.20& 0.22& 0.18&0.11   \\
       \hline\hline
       \end{tabular}
       \label{Tab:2}
\end{table}


\newpage
\section{Conclusion}\label{conclusions}

Two prototypes, four QM, eleven FM and three FS OMTs have been built and extensively tested from the electromagnetic and mechanical point of view. 
The design was optimized to 
meet the stringent requirements imposed by the scientific goal of the PLANCK mission. 
The very large bandwidth  requirements had to be met in a very small length, especially at 30 and 44 GHz. The orientation of the receiver ports at 30 and 44 GHz required a twist on the main arm, and reflections on both arms had to be of comparable extent. Isolation should be very large.
It was not possible to meet all the electromagnetic specifications, especially those concerning the RL, 
while coping with the mechanical constraints.
However, the  actual performances were
quite acceptable and in line with the overall LFI scientific requirements, with a simple and compact structure, suitable also for the higher frequency band, without posts or tuning elements.
During the QM/FM phase the transmission and reflection coefficient and the isolation between the two arms of the OMT have been measured over the widest possible frequency  band. Moreover, the agreement between measurements and simulations 
resulted good, giving confidence in the manufacturing process.  Although it was not possible to measure the OMT's performance at operational (cryogenic) temperature, for the 30 and 44 GHz channel the measured losses at room temperature were already compliant with the requirement at 20~K. At 70 GHz the losses at cryogenic temperature were estimated to be always within requirements. 
All the units showed high reproducibility of insertion and return loss. 
Significant differences were found in the cross polarization and isolation level that are much more sensitive to manufacturing tolerances. 
Finally, all FM OMTs were successfully vibrated for space qualification, after integration with the feed horns and after the reinforcement of the
flanges, satisfying  the entire electro-mechanical qualification process.

\begin{center}
 \textbf{acknowledgements}
\end{center}

    Planck is a project of the European Space Agency with instruments
funded by ESA member states, and with special contributions from Denmark
and NASA (USA). The Planck-LFI project is developed by an International
Consortium lead by Italy and involving Canada, Finland, Germany, Norway,
Spain, Switzerland, UK, USA. The Italian contribution to Planck is
supported by the Italian Space Agency (ASI).

\end{document}